\documentclass[manuscript]{aastex}

\usepackage{natbib}

\def \fermi {{\it Fermi}-LAT }

\def \mt {  }

\slugcomment{\bf \noindent
Submitted to \textit{ApJLetters}, 17 Aug., 2009; revised: 8 Jan.
2010}

\shorttitle{AGILE observations of the Supernova Remnant IC 443.}

\shortauthors{Tavani et al. 2009}

\begin{document}

\title{Direct evidence for hadronic cosmic-ray
acceleration \\in the Supernova Remnant IC 443 }

\bigskip

\author{
M.~Tavani\altaffilmark{1,2}, A.~Giuliani\altaffilmark{3},
A.~W.~Chen\altaffilmark{3,4},
 A.~Argan\altaffilmark{1},
G.~Barbiellini\altaffilmark{6},
A.~Bulgarelli\altaffilmark{5}, P.~Caraveo\altaffilmark{3},
 P.~W.~Cattaneo\altaffilmark{7},
 V.~Cocco\altaffilmark{1}, T.~Contessi\altaffilmark{3},
F.~D'Ammando\altaffilmark{1,2}, E.~Costa\altaffilmark{1}, G.~De
Paris\altaffilmark{1}, E.~Del Monte\altaffilmark{1},
G.~Di~Cocco\altaffilmark{5}, I.~Donnarumma\altaffilmark{1},
Y.~Evangelista\altaffilmark{1}, A.~Ferrari\altaffilmark{4,18}
M.~Feroci\altaffilmark{1},
F.~Fuschino\altaffilmark{5}, M.~Galli\altaffilmark{8},
F.~Gianotti\altaffilmark{5}, C.~Labanti\altaffilmark{5},
I.~Lapshov\altaffilmark{1}, F.~Lazzarotto\altaffilmark{1},
P.~Lipari\altaffilmark{9}, F.~Longo\altaffilmark{6},
M.~Marisaldi\altaffilmark{5},
 M.~Mastropietro\altaffilmark{10},
S.~Mereghetti\altaffilmark{3},
 E.~Morelli\altaffilmark{5}, E.~Moretti\altaffilmark{6},
A.~Morselli\altaffilmark{11}, L.~Pacciani\altaffilmark{1},
A.~Pellizzoni\altaffilmark{17}, F.~Perotti\altaffilmark{3},
G.~Piano\altaffilmark{1,2,11}, P.~Picozza\altaffilmark{2,11},
M.~Pilia\altaffilmark{20},
G.~Pucella\altaffilmark{13}, M.~Prest\altaffilmark{20},
M.~Rapisarda\altaffilmark{13}, A.~Rappoldi\altaffilmark{7},
E.~Scalise\altaffilmark{1}, A.~Rubini\altaffilmark{1},
S.~Sabatini\altaffilmark{2,11}, E.~Striani\altaffilmark{2,11},
P.~Soffitta\altaffilmark{1}, M.~Trifoglio\altaffilmark{5},
A.~Trois\altaffilmark{1}, E.~Vallazza\altaffilmark{6},
S.~Vercellone\altaffilmark{16}, V.~Vittorini\altaffilmark{1,2},
A.~Zambra\altaffilmark{3}, D.~Zanello\altaffilmark{9},
C.~Pittori\altaffilmark{14},
 F.~Verrecchia\altaffilmark{14}, P.~Santolamazza\altaffilmark{14},
P.~Giommi\altaffilmark{14}, S.~Colafrancesco\altaffilmark{14},
L.A.~Antonelli\altaffilmark{19},  L.~Salotti\altaffilmark{15}}

\altaffiltext{1} {INAF/IASF-Roma, I-00133 Roma, Italy}
\altaffiltext{2} {Dip. di Fisica, Univ. Tor Vergata, I-00133 Roma,
Italy} \altaffiltext{3} {INAF/IASF-Milano, I-20133 Milano, Italy}
\altaffiltext{4} {CIFS-Torino, I-10133 Torino, Italy}
\altaffiltext{5} {INAF/IASF-Bologna, I-40129 Bologna, Italy}
\altaffiltext{6} {Dip. Fisica and INFN Trieste, I-34127 Trieste,
Italy} \altaffiltext{7} {INFN-Pavia, I-27100 Pavia, Italy}
\altaffiltext{8} {ENEA-Bologna, I-40129 Bologna, Italy}
\altaffiltext{9} {INFN-Roma La Sapienza, I-00185 Roma, Italy}
\altaffiltext{10} {CNR-IMIP, Roma, Italy} \altaffiltext{11} {INFN
Roma Tor Vergata, I-00133 Roma, Italy} \altaffiltext{12} {Dip. di
Fisica, Univ. Dell'Insubria, I-22100 Como, Italy}
\altaffiltext{13} {ENEA Frascati,  I-00044 Frascati (Roma), Italy}
\altaffiltext{14} {ASI Science Data Center, I-00044
Frascati(Roma), Italy} \altaffiltext{15} {Agenzia Spaziale
Italiana, I-00198 Roma, Italy}
 \altaffiltext{16} {INAF/IASF Palermo, Italy}
 \altaffiltext{17} {INAF-Osserv.
Astron. di Cagliari,
I-09012 Capoterra, Italy} \altaffiltext{18} {Dip. Fisica,
Universit\'a di Torino, Turin, Italy} \altaffiltext{19}
{INAF-Osserv. Astron. di Roma, Monte Porzio Catone, Italy}
\altaffiltext{20} {Dip. Fisica, Universit\'a dell'Insubria,
I-22100 Como, Italy}

\begin{abstract}

The Supernova Remnant (SNR) IC~443 is an intermediate-age remnant
well known for its radio, optical, X-ray and gamma-ray energy
emissions. In this {\it Letter} we study the gamma-ray emission
above 100 MeV from IC~443 as obtained by the AGILE satellite. A
distinct pattern of diffuse
emission in the energy range
100 MeV-3 GeV is detected across the SNR with its prominent
maximum (source "A") localized in the Northeastern
shell {\mt with a flux $F = (47 \pm 10) \cdot 10^{-8} \, \rm
photons \, cm^{-2} \, s^{-1}$ above 100~MeV. }
 This location is the site of the strongest shock
interaction between the
SNR blast wave and the dense
circumstellar medium.
Source "A" is {\it not} coincident with the TeV source located 0.4
degree away and associated with a dense molecular cloud complex in
the SNR central region.
From our observations, and from the lack of detectable diffuse TeV
emission from its Northeastern rim, we demonstrate that electrons
cannot be the main emitters of gamma-rays in the range 0.1-10 GeV
at the site of the strongest SNR shock. The intensity, spectral
characteristics, and location of the most prominent gamma-ray
emission together with the
absence of co-spatial
detectable TeV emission are consistent only with a hadronic model
of cosmic-ray acceleration in the SNR. A high-density molecular
cloud (cloud "E") provides a remarkable "target" for nucleonic
interactions of accelerated hadrons: our results
show enhanced gamma-ray production near the molecular
cloud/shocked shell interaction site.
IC~443 provides the first unambiguous evidence of cosmic-ray
acceleration by SNRs.


\end{abstract}


    \section{Introduction}

Galactic cosmic-rays (CRs)
are believed to be accelerated above $10^{14}$-$10^{15}$~eV
energies by powerful supernovae in our Galaxy
(e.g., Shklovskii 1953; Ginzburg \& Syrovatskji 1964; Cesarsky
1980; Blandford \& Eichler 1987),
and gamma-rays above 70 MeV  are expected to provide the crucial
signature of hadronic acceleration. Shock-accelerated CRs
interacting with the gaseous surroundings of Supernova Remnants
(SNRs) produce gamma-rays by nucleon-nucleon interactions and
neutral pion decay. Indeed, EGRET observations of SNRs have
provided several important gamma-ray/SNR associations
\citep{sturner,esposito}. However, the complex morphology of SNRs
and the
EGRET angular resolution
prevented a definite resolution of this issue {\mt (e.g., Torres
et al. 2003)}. In recent years, TeV detections of SNRs provided
additional and very promising elements (e.g., Aharonian 2004;
Aharonian~et~al.~2007a,~2007b, 2008; Yamazaki~et~al.~2006;
Enomoto~et~al.~2006; Berezkho~\&~Voelk~2006;
Albert~et~al.~2007,~2008;~Acciari~et~al.~2009).
However,~the~hadronic~interpretation of these detections usually
requires a knowledge of
{\mt SNR } physical parameters or processes (e.g., electrons'
Bremsstrahlung/inverse Compton emission vs. hadronic pion
production, magnetic field strengths, the electron/proton number
ratio for GeV-TeV kinetic energies, etc.). {\mt Difficulties in
the interpretation of SNR gamma-ray and TeV data remain because of
our poor knowledge of SNR distances and ages, ambient gas density,
and problematic determinations of the non-thermal synchrotron and
high-energy emissions.}
%

\setcounter{footnote}{0}

 To { unambiguously} prove  the CR
acceleration by supernovae  we need SNRs for which we can {
demonstrate} that the ubiquitous and
co-accelerated
electrons do not dominate the observed gamma-ray and TeV emission.
 Given the
variety of SNRs and the complexity of interactions with their
environments, this task turned out to be very difficult to
accomplish\footnote{See also Butt 2009, for a recent review.}.

Current gamma-ray instrumentation can substantially improve this
picture. Detailed gamma-ray mapping of SNRs can provide a first
piece of evidence, i.e., confirming whether pion-generated
emission in the energy range 100 MeV-10 GeV is mostly concentrated
in sites where molecular clouds are strongly shocked by SNR
expanding shells.
As mentioned above, a second piece of evidence is required,
showing  that hadrons and not electrons are the main contributors
to the detected gamma-ray emission.  It turns out that IC~443
provides both pieces of evidence.

\section{The Supernova Remnant IC 443}

The intermediate age SNR IC 443 ($\tau \sim \,10$-20 thousand
years) is located near the  Galactic anticenter (l= 189.1, b =
+3.0), and
 is at a relatively close distance from the Earth ($\sim 1.5$
kpc) \citep{welsh}. It is one of the best studied SNRs because of
its complex structure and interaction with its gaseous
surroundings in the absence of strong diffuse Galactic emission.
Radio \citep{mufson,leahy}, 
 optical 
 \citep{fesen} and X-ray 
(e.g., Petre et al. 1988; Asaoka \& Aschenbach 1994; Kawasaki et
al. 2002; Troja et al. 2006)
 mapping of the SNR
show an asymmetric shape. The North-East rim expands in a
relatively dense environment
 with a shock velocity of $v_s \sim 65-100 \, \rm km
\, s^{-1}$ (for an average number density of the unperturbed
medium $n_1 \sim 10-100 \rm \, cm^{-3}$) \citep{fesen}, and the
South-West shell expands in a much more dilute medium ($n_2 \sim
1-10 \rm \, cm^{-3}$). In the middle of the SNR, a dense and very
massive ($\sim 1000 \, M_{\odot}$) molecular cloud complex in a
form of a torus or "ring" surrounds from the exterior the
expanding SN shell
(Dickman et al. 1992, Lee et al. 2008). Carbon monoxide (CO)
mapping of IC 443
shows several other smaller molecular clouds that interact with
the expanding shell.
In particular, the molecular cloud complex named "E" with a
projected size of $\sim 1$pc and mass estimate equals to $23 \,
M_{\odot}$ is the only prominent mass clump located just in front
of the expanding SN shell in the Northeastern rim
(Dickman et al. 1992). Furthermore, detailed mapping  of the J =
1-0 line of the formyl ion (HCO+, that traces compressed and
heated gas within the SNR) clearly shows that both the massive
"ring" and
cloud "E" are physically interacting with the SN ejecta. 
Given the morphology of the SNR and its molecular clouds
environment, IC 443 is therefore an ideal system to test the
hypothesis of hadronic CR acceleration in SNRs. It provides a
system which has all the required ingredients: a powerful SN of
total explosion energy $W \sim  10^{51}$ ergs, both dilute and
dense ambient gaseous environments surrounding the SN blast wave,
and a complex of molecular clouds physically interacting with the
SN shock.


\section{ AGILE Gamma-Ray Observations of IC 443}

The AGILE satellite has been operating since 2007 April 23,
 \citep{tavani-1}.
The AGILE scientific instrumentation is very compact and is based
on  two co-aligned imaging detectors operating in the energy
ranges 30 MeV - 30 GeV (GRID, Barbiellini et al. 2002, Prest et
al., 2003) and 18-60 keV (Super-AGILE, Feroci et al. 2007). An
anticoincidence system \citep{perotti}, a calorimeter
\citep{labanti}, and a data handling system \citep{argan} complete
the instrument. AGILE's performance is characterized by very large
fields of view (2.5 and 1 sr for the gamma-ray and hard X-ray
bands, respectively), optimal angular resolution and good
sensitivity (see Tavani et al. 2009 for details about the mission
and main instrument performance).
%
%

\begin{figure*}
%
 \begin{center}
 \includegraphics [height=10.cm]{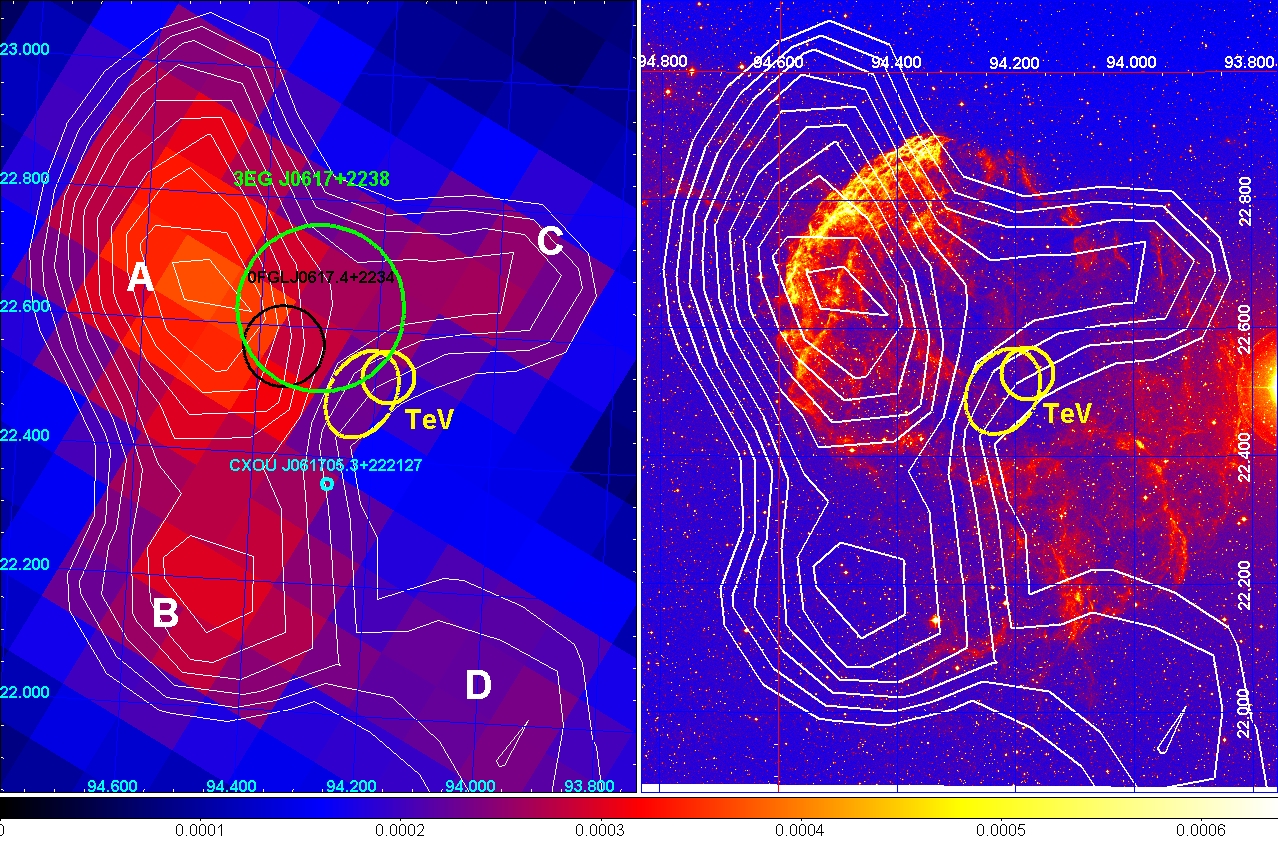}
 \caption{\small Gamma-ray intensity maps (J2000 R.A. and DEC coordinates)
  above 400 MeV of IC443 obtained by integrating all AGILE data.
 ({\it Left panel:}) gamma-ray intensity map above 400 MeV centered on
 IC~443. The color bar scale is in units of $\rm photons \, cm^{-2} \,
s^{-1} \, pixel^{-1}$. Pixel size is 0.1 degrees, and we used a
3-bin Gaussian smoothing. White contour levels of the gamma-ray
intensity start from 0.0002 and increase in steps of 0.0000173. We
also mark the positions and circular approximations of the 95\%
error boxes of the EGRET source 3EG J0617+2238 (green circle) and
\fermi 0FGL~J0617.4+2234 (black circle). The position of the TeV
source associated with IC~443
is marked by a yellow circle and ellipse that give the 95\%
confidence level error boxes determined by MAGIC \citep{albert}
and VERITAS \citep{acciari}, respectively (see also Butt 2009).
The X-ray source CXOU~J061705.3+222127 is marked by a cyan circle.
({\it Right panel:}) optical image of IC 443 (Palomar Digitized
Sky Survey) superimposed with the AGILE gamma-ray intensity
contours above 400 MeV (same as left panel). The position of the
TeV source is marked by the yellow circle (MAGIC) and ellipse
(VERITAS).
 The angular distance between
the centroids of the gamma-ray dominant source in the Northeastern
rim and the TeV source is $\sim 0.4$ degrees. The TeV source is
located in the central part of the IC 443 at the location
coincident with a massive molecular cloud complex \citep{dickman}.
 At the estimated distance of 1.5 kpc, 1 arcmin subtends a
distance of 0.44 pc. }
\label{fig-1}
\end{center}
\end{figure*}

During the first two years of operations AGILE observed several
times the Galactic anticenter region.
A total observing time of approximately 1 Msec was
accumulated by AGILE,
 on IC~443,
and a high-resolution mapping and spectral analysis in the energy
range 100 MeV - 20 GeV were obtained using using standard AGILE
gamma-ray selection procedures.  Positional astrometry has been
carefully checked by comparing the IC~443 gamma-ray emission with
 the nearby Crab and Geminga pulsars.
The region is not substantially affected by the weak diffuse
gamma-ray Galactic emission that is properly taken into account in
our analysis \cite{giuliani}.

Fig.~1 shows the result of the AGILE detailed gamma-ray mapping
(above 400 MeV) superimposed to an optical map of the Northeastern
rim of IC 443. Diffuse gamma-ray emission with significant
enhancements is detected across IC~443 in a pattern that closely
resembles the SNR outer shell configuration. A most prominent
gamma-ray enhancement (that we label "source A" in Fig.~1) is
clearly detected
in the Northeastern region of IC~443 at a location coincident with
the most active expanding SN blast wave. We also detect several
other gamma-ray enhancements that follow a pattern surrounding the
outer regions of the SNR shell. We label the remaining
enhancements "B", "C", and "D" (see Fig.~1). The gamma-ray pattern
of emission from IC 443 does not show any sign of significant
variability. In this paper, we concentrate our analysis on the
Northeastern rim of IC~443 and on  source "A", postponing a more
detailed study of the SNR to a forthcoming publication.

Source "A" is detected at 11 sigma level at the Galactic
coordinate location (l,b): $(189.08, 3.28) \pm 0.17 \, \rm (stat.)
\, \pm \, 0.1 \, (syst.)$, with a flux $F = (47 \pm 10) \cdot
10^{-8} \, \rm photons \, cm^{-2} \, s^{-1}$ above 100~MeV. We
notice that the
position of source "A" is consistent 
with that of the only prominent molecular cloud in the
Northeastern region, i.e., cloud "E" of Dickman et al. 2002 which
is in physical interaction with the expanding shell. Clearly
marked in Fig.~1 is the relatively small error
box of the most prominent TeV source detected by MAGIC 
\citep{albert} and
VERITAS \citep{acciari}. 
The pulsar wind nebula
CXOU J061705.3-222127 \citep{gaensler} is  not
associated with any prominent gamma-ray emission.

Within the statistical and systematic uncertainties, the source
"A" location is consistent with the EGRET (3EG~J10617+2238,
Hartman et al. 1999),
AGILE
(AGL~J0617+2236, Pittori et al. 2009) and \fermi
(0FGL~J0617.4+2234, Abdo et al. 2009) sources.
While the EGRET source appeared to be marginally compatible with
the position of the TeV source, the refined AGILE location is
inconsistent at more than the 99 \% confidence level with that of
the TeV source.
The location of the most intense TeV emission is indeed
concentrated in the central part of IC 443 in apparent
superposition with the massive molecular cloud "ring". The angular
distance between source "A" and the TeV source centroid is more
than 0.4 degrees, i.e., significantly larger than the location
accuracies of AGILE (0.1 degree for this integration) and of MAGIC
and VERITAS. We conclude that the positional difference between
the gamma-ray source "A" and the TeV source is significant, and
reflects the difference in the physical locations of the dominant
0.1-10 GeV and TeV emissions, respectively. We infer a very
important fact for the physical interpretation of our
observations: the most prominent 100 MeV-10 GeV emission is not
co-spatial with the TeV source. Furthermore, AGILE detects weak
and  diffuse gamma-ray emission from the central location of
IC~443 in coincidence with the TeV source and the centroid of the
massive molecular "ring".

\section{Discussion}

Let us denote by  $\chi_{e,p}$ the ratio between the electron and
hadron number density normalizations (at momentum $p = $~1~GeV/c).
Electrons
can emit {\mt high-energy photons} by Bremsstrahlung and inverse
Compton (IC) scattering. Emission by these two processes is
unavoidably linked, in the sense that if an electron
Bremsstrahlung contribution emerges as a prominent spectral
component, a strong IC contribution is predicted in the TeV range.
The same electrons emitting Bremsstrahlung gamma-rays (possibly
enhanced by a dense cloud) also scatter unavoidably the ambient
optical/IR photon bath of the SNR/interstellar medium and the
cosmic microwave background (CMB) photons. A general conclusion
can be deduced for a typical SNR: a ratio $\chi_{e,p} \sim 1$
implies  {\it co-spatial} ~100 MeV and ~TeV emissions in SNRs. A
detailed modelling of electron (and proton) emissivities of IC~443
indeed  predicts
  that for an ideal matter density of  $n_o \sim 1
\rm \; cm^{-3}$ the IC contribution to the spectral power ($\nu
F_{\nu}$)
 dominates by a factor of $\sim 10$ over the Bremsstrahlung contribution
 for photon
energies between 100 MeV and 1 GeV  (Gaisser, Protheroe \& Stanev
1998, GPS98). For higher energies, the IC dominance is even
larger, leading to a predicted IC power at 300 GeV larger that a
factor of $\sim 100$ than the Bremsstrahlung spectral power. If we
apply these predictions to the specific case of the IC~443
Northeastern rim characterized by an external medium density $n_1
\sim 10-100 \rm \; cm^{-3}$, we expect the Bremsstrahlung power
$(\nu F_{\nu})_B$ near 0.1-1 GeV and the IC power  $(\nu
F_{\nu})_{IC}$ near 0.3-1 TeV to be approximately equal. This
implies that a model with $\chi_{e,p} \sim 1$ predicts the
electron-driven TeV emission to be co-spatial with the
Northeastern rim, and in particular to be coincident with the
0.1-1 GeV emission. Since this is not observed, we conclude that
the SN shock must be characterized by a ratio  $\chi_{e,p}$
substantially less than unity. We note that a value $\chi_{e,p}
\sim 0.01$ has been indirectly deduced
for IC~443 (GPS98) and other SNRs (e.g., Aharonian 2004), and in
general agrees with what observed in direct CR measurements.

The overall pattern of gamma-ray emission from IC~443, combined
with the absence of detectable co-spatial TeV emission both near
source "A" and over the whole semispheric shocked Northeastern
rim, has far reaching consequences. Fig.~2 shows the results of
two leading models of gamma-ray emissivity produced by protons and
electrons near source "A". {\mt We follow the treatment of
Drury et al. 1994, and Berezhko \& V\"olk 1997 for the
normalization of the expected gamma-ray emission from accelerated
hadronic CRs in terms of parameters of the late SNR evolutionary
phase giving $E_{\gamma}^2 \, d\, N_{\gamma}/d \, E_{\gamma}
\simeq 5 \times 10^{-11} {\rm \, (TeV \, cm^{-2} \, s^{-1})} \,
(E_{SN}/10^{51} \, {\rm erg}) \, (d/1 \,  {\rm kpc})^{-2} \,  n_1
\, f$, where $E_{SN}$ is the total supernova explosion energy, $d$
the source distance, and $f$ a geometrical factor. }
For simplicity, we assume that hadrons and electrons are
accelerated at the Northeastern shock with the same power-law
energy index, i.e., $N(E) = k \, E^{-\alpha}$. Models without
energy cutoffs below 10 TeV for both electrons ($E_{c,e}$) and
protons ($E_{c,p}$) contradict the observations. We consider then
two models\footnote{{\mt In order to display the dynamic range of
the most important parameter influencing the shape of the
spectrum, we consider the values $\alpha = 2.1$ (a realistic
case), and $\alpha = 2.7$ (a case with no spatial diffusion).}}
{\mt for source "A"} as constrained by the data {\mt with
$E_{SN}=10^{51} \, \rm erg$, $d = 1.5$~kpc, and $f \simeq 0.01$.}
Model 1 is characterized by $\chi_{e,p} = 0.03$, $\alpha = 2.7$,
$n_1 = 10^3 \rm \; cm^{-3}$, $E_{c,e} = 10$~ TeV, $E_{c,p} = 200
$~GeV. Model~2 has $\chi_{e,p} = 0.01$, $\alpha = 2.1$, $n_1 =
10^3 \rm \; cm^{-3}$, $E_{c,e} = 100 $~GeV, $E_{c,p} = 100 $~GeV.
Fig.~2 clearly shows that the hadron-produced gamma-ray emissivity
has to be suppressed between 10 and 100 GeV. The electron
Bremsstrahlung and IC contributions are also shown; both have to
be consistent with the absence of co-spatial gamma-ray and TeV
emission in the Northeastern rim
satisfying the observational constraint $\nu F_{\nu}({\rm
100MeV-10GeV}) \gtrsim 100 \; \nu F_{\nu}({\rm 0.2-1 TeV})$. {\mt
We note that a value of $\chi_{e,p} \sim 0.01$ is consistent with
the radio synchrotron emission observed in the Northeastern part
of the IC~443 shell \citep{erickson}, for an average local
magnetic field $B \sim 10^{-5}\,$G.}

\begin{figure}
\begin{center}
\vspace*{-.4cm} \hspace*{-0.8cm}
\includegraphics [height=11.cm]{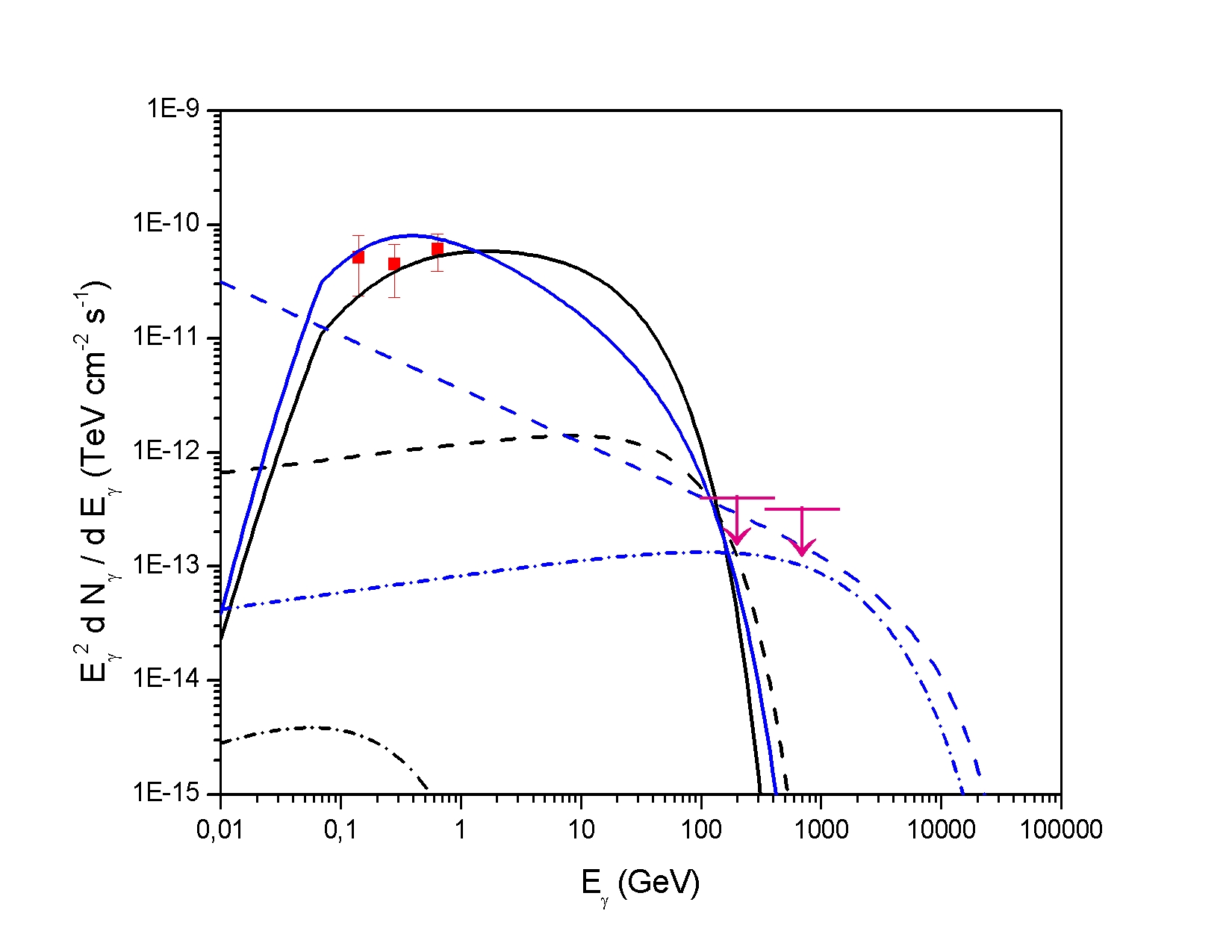}
\vspace*{-0.4cm}
 \caption{ \small
The AGILE gamma-ray spectrum (between 100 MeV and 1 GeV) and the
TeV upper limits of the region centered at the source "A"
of IC 443  compared with two different models for the gamma-ray
emissivities. \textit{Solid curves:} hadronic production of
gamma-rays by neutral pion decay (blue curve: model 1, black
curve: model 2). \textit{Dashed curves:} Bremsstrahlung emission
by electrons (blue curve: model 1, black curve: model 2).
\textit{Short-dash-dot curves:} electron IC emission on the cosmic
microwave background  (blue curve: model 1, black curve: model 2).
The MAGIC upper limits are assumed to have values 3 times smaller
than the reported flux levels 
of the TeV source at the center of IC 443 \citep{albert,acciari}.
The AGILE spectral data agree with those reported by EGRET
\citep{esposito}. }
 \vspace{-0.1cm} \label{fig-2}
 \end{center}
 \end{figure}


\section{Conclusions}

We obtain a satisfactory physical picture of IC~443 with very
important implications for the problem of the cosmic-ray
production by SNRs. Diffuse gamma-ray emission of hadronic origin
is remarkably distributed in coincidence with  the IC~443 outer
shock. A clear gamma-ray enhancement is detected (source "A") in
the Northeastern rim and close to the molecular cloud "E". This
cloud provides an ideal target for nucleonic interactions of
accelerated hadrons (protons and ions) with subsequent gamma-ray
emission by neutral pion decay. Cloud "E" has a substantial target
matter density for efficient hadronic interaction with a timescale
$\tau_{pp \longrightarrow \pi^o} \sim (5 \cdot 10^4 \; {\rm yrs})
\, (n/10^3 \; \rm cm^{-3})^{-1}$. This timescale turns out to be
comparable with the age of the remnant, and implies a high
efficiency of gamma-ray production by the SN blast wave-cloud
interaction.
By assuming an average particle kinetic energy of order of 0.1$-$1
TeV, our gamma-ray observations of IC~443 and the proposed
physical interpretation are in agreement with the total energy in
cosmic-ray particles being 1-10\% of the total estimated SN
explosion energy of about $10^{51}$ ergs.

At the source "A" site,
both the proton and electron distributions have an effective
energy cutoff of order of 100 GeV.  We note that several models
 of particle acceleration in high-density regions close
to the acceleration site consider the possibility of effective
energy cutoffs of order 0.1-1 TeV (e.g., Drury et al. 1996, Baring
et al. 1999, Malkov et al. 2002).
These effective cutoffs can be caused by a combination of
inefficient acceleration at larger energies and particle
diffusion. {\mt The CR maximum kinetic energy $E_m$ produced by
shock acceleration can be estimated as $D(E_m)= R_s \, V_s / k'$,
where $D(E)$ is the particle spatial diffusion coefficient, $k'
\simeq 30$ for the late evolutionary phase, and $R_s$ and $V_s$
are the shock radius and speed, respectively (e.g., Berezhko
1996). An upper limit for $E_m$ applicable for source "A" of
IC~443  can be obtained with a Bohm approximation for $D$, and
$R_s = 1 \,$~pc, $V_s \sim 10^8 \, \rm cm^{-2} \, s^{-1}$, and $B
\sim 5 \times 10^{-6} \,$G. We obtain $E_m \sim 1 \,$TeV,
indicating that the active SNR shocked region currently contains
only relatively low kinetic energy particles with the highest
energy CRs diffused away a long ago.}
 Our observations of the intermediate-age
IC~443
support this important aspect of the particle acceleration
mechanism.

It is also interesting  to address the lack of prominent gamma-ray
emission in coincidence with the TeV source in the high-density
medium at the center of the SNR. A possible explanation is based
on  a combination of  energy dependent particle diffusion and
nucleonic interactions in the very dense central medium reached by
the SN shock \citep{torres-2}.
In any case, the absence of detectable TeV emission from most of
the SNR argues for a sub-dominant electron contribution to the
emission.

 We presented here two specific emission models in agreement with the
 high-energy observations and yet predicting different fluxes in the TeV range.
 Future deep TeV observations of IC~443 may reveal a weak TeV emission associated
 with the Northeastern rim and the source "A" region.
 Fig.~2 provides an example of how these observations may help in further constraining
 the theoretical models.

IC 443 turns out to be the first SNR clearly providing evidence
for hadronic cosmic-ray acceleration and interaction with its
gaseous surroundings. A leptonic model of emission is in
contradiction with the combined gamma-ray/TeV observations of the
Northeastern rim of IC~443 and in particular of the
source "A" region. A hadronic model of emission agrees in a
natural way with our observations, and confirms the hypothesis
that supernova blast waves can efficiently accelerate protons and
possibly other ions in our Galaxy.

\acknowledgments {\mt We thank an anonymous referee for his/her
comments that led to an improvement of the discussion of our
results.} The AGILE Mission is funded by the Italian Space Agency
(ASI) with scientific and programmatic participation by the
Italian Institute of Astrophysics (INAF), and the Italian
Institute of Nuclear Physics (INFN). This investigation was
carried out with partial support from the ASI contract
n.~I/089/06/2.


{}


\begin{thebibliography}{}







\bibitem[Abdo et al. 2009]{abdo}  Abdo, A.A., et al.
2009, http://fermi.gsfc.nasa.gov

\bibitem[Acciari et al. 2009]{acciari}
Acciari, V.A.,  et al., 2009, ApJL, in press (arXiv:0905.3291v1
[astro-ph])

\bibitem[Aharonian 2004]{aharonian}
Aharonian, F.A., 2004, \textit{Very High Energy Cosmic Gamma
Radiation} (Singapore: World Scientific)



\bibitem[Aharonian et al. 2007a]{aharonian-2}
Aharonian, F.A., et al., 2007a, A\&A, 464, 235

\bibitem[Aharonian et al. 2007b]{aharonian-3}
Aharonian, F.A., et al.,  2007b, ApJ, 661, 236

\bibitem[Aharonian et al. 2008]{aharonian-4}
Aharonian, F.A., et al., 2008, A\&A, 481, 401

\bibitem[Albert et al. 2007]{albert}
Albert, J., et al., 2007, ApJ, 664, L87

\bibitem[Albert et al. 2008]{albert-2}
Albert, J., et al., 2008, ApJ, 674, 1037


\bibitem[Argan et al. 2004]{argan}
Argan, A., et al., 2004, Proc. IEEE-NSS, 1, 371

\bibitem[Asaoka \& Aschenbach 1994]{asaoka}
Asaoka, I., \&   Aschenbach, B., 1994, A\&A, 284, 573

\bibitem[Barbiellini et al. 2002]{barbiellini}
Barbiellini G., et al.,
2002, NIM A, 490, 146.


\bibitem[Baring et al. 1999]{baring}
Baring, M.G., Ellison, D.C., Reynolds, S.P., Grenier, I.A., P.
Goret, P., 1999, ApJ, 513, 311


 \bibitem[Berezhko 1996]{berezhko-96}
 Berezhko, E.G., 1996, Astropart. Phys., 5, 367


 \bibitem[Berezhko \& Voelk 1997]{berezhko}
 Berezhko, E.G., \& Voelk, H.J., 1997, Astropart. Phys., 14, 183


\bibitem[Berezhko \& Voelk 2006]{berezhko-2}
Berezhko, E.G., \& Voelk, H.J., 2006, A\&A, 451, 981




 \bibitem[Blandford \& Eichler 1987]{blandford} Blandford, R., \&
 Eichler, D., 1987, Phys. Rep., 154, 1


\bibitem[Butt 2009]{butt}
Butt, Y., 2009, Nature, 460, 701

\bibitem[Cesarsky 1980]{cesarsky}
Cesarsky, C., 1980, Ann. Rev. Astron. \& Astrophys., 18, 289


\bibitem[Dickman et al. 1992]{dickman}
Dickman, R.L.,  Snell, R.L., Ziurys, L.M., \&   Huang, H.L., 1992,
ApJ, 400, 203


 \bibitem[Drury, Aharonian \& Voelk 1994]{drury-2}
 Drury, L.O'C., Aharonian, F.A. \& Voelk, H.J., 1994, A\&A, 287,
 959


 \bibitem[Drury, Duffy \& Kirk 1996]{drury}
 Drury, L.O'C., Duffy, P., \& Kirk, J.G., 1996, A\&A, 309, 1002

\bibitem[Enomoto et al. 2006]{enomoto}
Enomoto, R., et al., 2006, ApJ, 652, 1268

\bibitem[Erickson \& Mahoney 1985]{erickson}
Erickson, W.C. \& Mahoney, M.J., 1985, ApJ, 290, 596

\bibitem[Esposito et al. 1996]{esposito} Esposito, J.A.,  Hunter, S.D., Kanbach, G.,
 Sreekumar, P., 1996, ApJ, 461, 820

 \bibitem[Feroci et al. 2007]{feroci}
Feroci M., et al., 2007, NIM A, 581, 728


\bibitem[Fesen \& Kirshner 1980]{fesen} Fesen, R.A., \& Kirshner, R.P., 1980, ApJ, 242,
1023

\bibitem[Gaensler et al. 2006]{gaensler}
Gaensler, B.M., Chattereje, S., Slane, P.O., van der Swaluw, E.,
Camilo, F., Hughes, J.P., 2006, ApJ, 648, 1037

\bibitem[Gaisser, Protheroe \& Stanev 1998]{gaisser}
Gaisser, T.K., Protheroe, R.J., \& Stanev, T., 1998, ApJ, 492, 219

\bibitem[Giuliani et al. 2004]{giuliani}
Giuliani, A., et al.,
2004, MmSAI,  5, 135

\bibitem[Ginzburg \& Syrovatskji 1964]{ginzburg}
 Ginzburg, V.L. \& Syrovatskji, S.I., 1964, \textit{The Origin of Cosmic Rays}
(Pergamon Press, New York, 1964)

\bibitem[Hartman et al. 1999]{hartman}
Hartman, R.C., et al., 1999, ApJ Suppl. Series, 123, 79




\bibitem[Kawasaki et al. 2002]{kawasaki}
Kawasaki, M.T., et al., 2002, ApJ, 572, 897

\bibitem[Yamazaki et al. 2006]{yamazaki}
Yamazaki, R., et al., 2006, MNRAS, 371, 1975

\bibitem[Labanti et al. 2006]{labanti} Labanti, C. et al., 2006, proc SPIE, 6266,
62663.

\bibitem[Lee et al. 2008]{lee}
Lee, J.-J., et al., 2008, AJ, 135, 796

\bibitem[Lehay 2004]{leahy} Leahy, D.A., 2004,  AJ, 127, 2277

\bibitem[Malkov, Diamond \& Jones 2002]{malkov}
Malkov, M.A., Diamond, P.H., Jones, T.W., 2002, ApJ, 571, 856

\bibitem[Mufson et al. 1986]{mufson} Mufson, S.L., et al., 1986, AJ, 92, 1349

\bibitem[Perotti et al. 2006]{perotti}
Perotti, F., et al., 2006, NIM A, 556, 228


\bibitem[Petre et al. 1988]{petre}
Petre, R.,  Szymkowiak, A.E.,   Seward, F.D.,  Willingale, R.,
1988, ApJ, 335, 215

\bibitem[Pittori et al. 2009]{pittori} Pittori, C., e al., 2009, submitted to
A\&A, http://agile.asdc.asi.it


\bibitem[Prest et al. 2003]{prest} Prest M., et al., 2003, NIM A, 501, 280

\bibitem[Shklovskii 1953]{shklovskii}
Shklovskii, I.S., 1953, Dokl. Akad. Nauk. SSSR, 91, 475

\bibitem[Sturner \& Dermer 1995]{sturner}
Sturner, S.J., \&  Dermer, C.D., 1995, Astr. \& Astrophys., 293,
17


\bibitem[Tavani et al. 2009]{tavani-1} Tavani, M., et al. 2009,
A\&A, 502, 995


 \bibitem[Torres et al. 2003]{torres}
 Torres, D.F., Romero, G.E.,  Dame, T.M., Combi, J.A.,  Butt, Y.M.,
 2003, Physics Rep., 382, 303



\bibitem[Torres et al. 2008]{torres-2}
Torres, D.F., Marrero, A.Y., \&  Cea del Pozo, E., 2008 MNRAS,
387, L59

\bibitem[Troja et al. 2006]{troja}
Troja, E., Bocchino, F., \& Reale, F., 2006, ApJ, 649, 258


\bibitem[Welsh \& Sallmen 2003]{welsh}
Welsch, B.Y., \&  Sallmen, S., 2003, A\&A, 408, 545



\end{thebibliography}
\end{document}